\documentclass[10pt, conference, compsocconf]{IEEEtran}
\usepackage[utf8]{inputenc}
\usepackage[T1]{fontenc}
\usepackage{graphicx}
\usepackage{grffile}
\usepackage{longtable}
\usepackage{wrapfig}
\usepackage{rotating}
\usepackage[normalem]{ulem}
\usepackage{amsmath}
\usepackage{textcomp}
\usepackage{amssymb}
\usepackage{siunitx}
\usepackage{capt-of}
\newcommand{\url}[1]{\texttt{#1}}
\usepackage{listings}

\usepackage{mathtools}
\usepackage[linesnumbered,ruled,vlined]{algorithm2e}

\author{
Ivan Porres,  Hergys Rexha, Sébastien Lafond\\
Faculty of Science and Engineering, \\ Åbo Akademi University \\ Turku, Finland \\
\texttt{name.surname@abo.fi}
}
\date{\today}
\title{Online GANs for Automatic Performance Testing}

\begin{document}

\maketitle
\begin{abstract}
In this paper we present a novel algorithm for automatic performance testing that uses an online variant of the Generative Adversarial Network (GAN) to optimize the test generation process. The objective of the proposed approach is to generate,  for a given test budget,  a test suite containing a high number of tests revealing performance defects. This is achieved using a GAN to generate the tests and predict their outcome. This GAN is trained online while generating and executing the tests. The proposed approach does not require a prior training set or model of the system under test.  We provide an initial evaluation the algorithm using an example test system,  and compare the obtained results with other possible approaches.

We consider that the presented algorithm serves as a proof of concept and we hope that it can spark a research discussion on the application of GANs to test generation.
 
\end{abstract}

\section{Introduction}
\label{sec:intro}

In this paper we propose a new algorithm for performance testing of computing systems that uses a Generative Adversarial Network as the main mechanism for test generation.

Performance testing is the process of identifying how a system performs in terms of different key performance indicators under varying usage. The end goal of performance testing is to find performance defects~\cite{subraya2000object, WEYUKER2000} by creating and executing tests against a system and observing its performance. In most cases, executing a test is an expensive operation and, therefore, an automated testing system should strive to execute tests only if they have a high likelihood to reveal a performance defect.

The new algorithm presented here has been developed on the basis of our previous work published in~\cite{DBLP:conf/icst/PorresARLT20} and~\cite{DBLP:conf/euromicro/PorresAL20}. These previous algorithms use random sampling to generate tests and a deep neural network as a test discriminator to decide if a test should be executed or not against the system under test. In~\cite{DBLP:conf/icst/PorresARLT20}, we assumed that there is a dichotomous oracle where the outcome of a test may only be pass or fail and used a supervised learning approach. In~\cite{DBLP:conf/euromicro/PorresAL20}, we extended our previous algorithm to support test outcomes based on real numbers and optimization of the test search based on real values.  In this article, we apply the concept of Generative Adversarial Networks (GAN)~\cite{DBLP:journals/corr/GoodfellowPMXWOCB14} to system testing. For this, we use two neural networks in the same algorithm, a generator network that proposes test candidates and a discriminator network that rejects these candidates if it predicts a negative outcome. 

GANs have been used successfully in many applications, specially within the domain of image generation. This is due to the fact that they have many benefits with respect to other generative methods~\cite{DBLP:journals/corr/Goodfellow17}. However, it seems that the use of GANs in Software Engineering applications is not common and, to our knowledge, this is the first work that uses a GAN for automated test generation. 

We proceed as follows: We present in Section II a description of the problem of automatic performance testing, as well as previous work related to this problem and the algorithm presented here. Section III contains the main contribution of the article, the presentation of an online GAN algorithm for performance test generation.  We evaluate the behavior of this algorithm experimentally using an example test system in Section IV. Finally, we discuss the main article contributions and concluding remarks in Section V.

\section{The Automatic Performance Testing Problem}

In this section we introduce the problem of automatic performance testing and related work to it. We focus on approaches that do not require a previous model or knowledge of the system under test.

We model the system under test as a function $S$, that accepts inputs from a nonempty set $I$. We also assume that there is a performance instrument $m$ that for a given system output $S(i), i \in I$, provides a measurement $m(S(i))$ corresponding to the performance of the system for the input.  Based on this, the  overall goal of performance testing is to find what inputs yield a performance   greater than a given threshold $p_m$: 

\[I_p=\{ i \in I : m(S(i))  \geq p_m\}\] 

We call $I_p$ the set of positive tests of the system, i.e., the system inputs that reveal a performance issue. Likewise, we can define $I_n=I-I_p$ as the set of negative inputs. 

In most cases, the set of possible inputs will be large and measuring the system performance for a given input will be an expensive operation. As a consequence, exhaustive exploration of the input set is not feasible in practice. Instead, we assume that there is a budget $n$ that limits the maximum number of test executions and corresponding performance measurements. In this case, for a given test budget $0<n \leq |I|$, we are interested in finding a best test suite $B$ such

 \[B(S,I,m,p_m)= \underset{T \subseteq I, |T|=n}{\arg\max} (|\{ t \in T : m(S(t)) \geq p_m\}|)\]

\subsection{Automatic Performance Testing as a Search Based Software Engineering Problem}

The previous definition formulates performance testing as an optimization problem. This should not be a surprise since  software testing often represents a trade-off in the quality-cost-time triangle. For example, the design of a unit test suite is a trade-off between source code coverage and execution time. 

These optimization problems are difficult to solve by exact algorithms due to their complexity and therefore there is a need for  heuristics that can provide near optimal solutions efficiently. For this reason, Search Based Software Engineering (SBSE) tackles the study and design of efficient solutions to these optimization problems in the context of Software Engineering applications.

A comprehensive survey on SBSE was published by Harman in 2012 \cite{DBLP:journals/csur/HarmanMZ12} and a survey on search-based testing was published by Ali~\cite{DBLP:journals/tse/AliBHP10}. SBSE is based on the more general field of search-based optimization (SBO) and the Handbook of Metaheuristics provides a comprehensive reference to the field \cite{10.5555/1941310}. However, generic SBSE algorithms  are hampered by two important issues that apply to existing metaheuristics. 

The first issue is revealed by the \emph{No Free Lunch} theorem for optimization~\cite{DBLP:journals/tec/DolpertM97}. It states that all metaheuristics, including random search, perform equally when considering all possible optimization problems. In practice, this means that in order to create a new SBSE solution, the chosen metaheuristic must be customized and tuned for each specific application.

The second issue lies in how to evaluate the fitness of solution candidate quickly. Unfortunately, the fitness function is often  computationally expensive in many SBSE applications~\cite{DBLP:journals/corr/NairYM17}. For example, in the case of software test generation, it may require to actually execute the test. As a consequence, the performance of a SBSE algorithm  is limited in practice by the cost of evaluating the fitness function.

\subsection{Machine Learning for SBSE}

Machine learning has found many applications within Software Engineering~\cite{DBLP:conf/iceis/BorgesCRP20} and we consider that it can help addressing the two issues mentioned above and become the basis for competitive automatic test generation algorithms.

To address the first issue, different authors have proposed the use of machine learning to drive the search of the solution space. Ahmad presented a reinforcement learning algorithm for performance testing in~\cite{DBLP:journals/access/AhmadATDP20}. Also, Nareyek proposed to use reinforcement learning as the high level hyper-heuristic~\cite{Nareyek2004}, an idea that has been refined by others, including the recent work by Choong~\cite{DBLP:journals/isci/ChoongWL18}. In this way, the algorithm can learn what low-level heuristic should be applied at each step and perform better over a larger set of problems.

For the second issue,  we can replace an expensive fitness function with a surrogate model. A survey on surrogate models for computationally expensive black-box functions can be found in~\cite{DBLP:conf/wcgo/Regis19}. The challenge is that choosing  a suitable approach to create the  surrogate model requires exploring the solution space, and this is precisely the operation that we try to avoid in order to save time. It is however possible to use learning to create a surrogate model online, while performing the search.  Nair has used a decision tree learner in discrete problems~\cite{DBLP:journals/corr/NairYM17} while we have used deep neural  networks to model  continuous functions~\cite{DBLP:conf/icst/PorresARLT20, SEAA2020}.


These recent advances in the use of learning to improve test search and surrogate model generation raise the question if we can combine both tasks in one single learning algorithm. The key observation is that both tasks, generation and surrogate model learning, must explore the same search space. 

We consider that such approach would be similar to a GAN. In a GAN, there is a deep neural network working as a generator and another network working as a discriminator. The generator  is in charge to produce new elements while the discriminator is in charge to determine if a given element is produced by the generator or it comes from a training set used as ground truth. The goal is to train the generator so it produces elements that  the discriminator cannot  distinguish from the training set.

Creating a new GAN on the existence of a training set, but this is not usually the case for automatic testing. Otherwise we could use that training set as an actual test suite. For this reason, we  adapt in the next section the concept of GANs to work with online supervised learning, without a prior training set.

\section{An Online GAN test generation method}

\begin{figure}[h!]
\begin{center}
\includegraphics[scale=0.35]{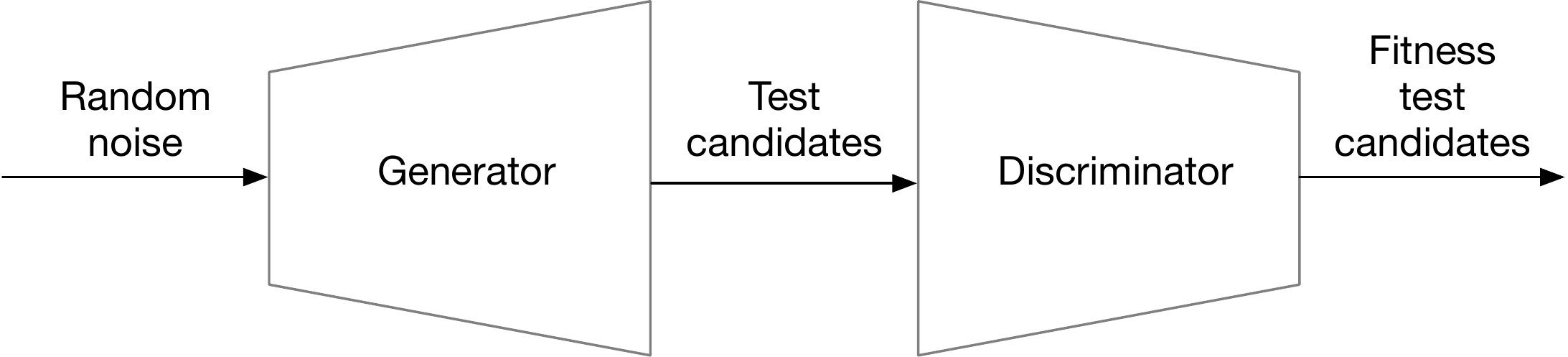}
\caption{Online GAN for Test Generation}\label{f:gan1}
\end{center}
\end{figure}

This section presents the main algorithm  proposed in this article: an online GAN test generator. The algorithm uses three components: the generator, the discriminator, and a system under test (SUT). The intuition behind  it is that the generator  explores the input space and produces candidate tests, while the discriminator evaluates the candidate tests and predicts their fitness. Figure~\ref{f:gan1} illustrates the GAN approach.

\begin{figure*}
\begin{center}
\includegraphics[scale=0.35]{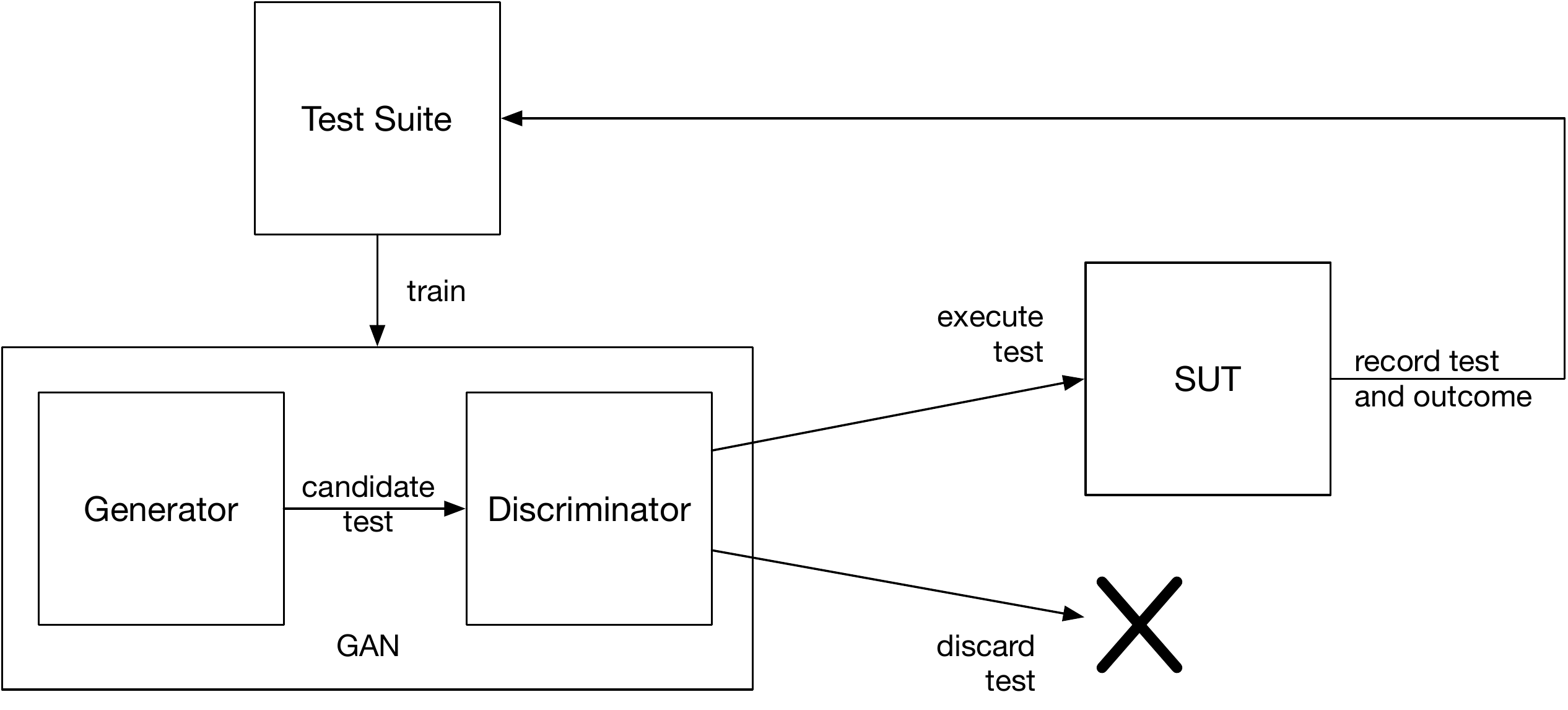}
\caption{Online GAN Test Generation System}\label{f:gan}
\end{center}
\end{figure*}

The predicted fitness can be used to reject the test candidates that may be negative tests. The candidate tests with a high fitness according to the discriminator are finally evaluated by the SUT (Figure~\ref{f:gan}), that establishes the fitness or ground truth for the test. We assume that evaluating a test by the SUT is a timewise expensive operation.

Without any prior data, the generator behaves like a random sampling algorithm, choosing tests from the input space at random. Likewise, the discriminator assigns a random fitness to each  test candidate and behaves as a coin flip to decide if a test candidate is worth being executed in the system under test. 

However, their performance can improve iteratively by learning online, while searching for new positive tests.  Similar to a traditional GAN,  the generator and the discriminator are in a game: The generator is trained to avoid being rejected by the discriminator while the discriminator is trained to avoid being fooled by the generator with a sub-optimal test candidate. After each  iteration, both the generator and discriminator improve in their tasks and become better at producing more relevant tests based on the actual measurements against the system under test.

We propose to use a deep neural network (DNN) as model for the generator and discriminator. We already explored the use of DNN as a discriminator in~\cite{DBLP:conf/icst/PorresARLT20}. However, other authors propose to use different models such as a decision-tree based optimizer, as published by of Nair, et al~\cite{DBLP:journals/corr/NairYM17}.

\subsection{The Online-GAN algorithm}

The online GAN algorithm is presented in detail in Algorithm~\ref{alg2}. Here, the function \emph{generate(NN,S,n)} uses the network \emph{NN} to produce a subset of $S$ with cardinality $n$, while the function \emph{predict(NN,d)} returns the prediction for the input $d$ and the network \emph{NN}.

\begin{algorithm}
\SetAlgoLined
\textbf{input} set of inputs I, integer budget \\
\textbf{requires} budget $\leq$ |I| \\
T := $\emptyset$\;
GAN := new model with generator GN and discriminator DN\;
 \While{\normalfont{|T|<budget}}{
   target := $1$\; 
   \Repeat{\normalfont{predict(DN,t) $\geq$ target}}{
   target := target*$treducer$; \\
   t:= generate(GN,I-[T]$_1$); \\
    }
   result := measure(execute(t))\;
   T := T $\cup$ \{(t,result)\}\;
   train(GAN, T)
  }
\textbf{result} test suite T
 \caption{Online GAN test generation algorithm}\label{alg2}
\end{algorithm}

The algorithm works as follows. Lines 3 and 4 initialize the test suite (T) to an empty set, and the GAN network to an untrained model. Lines 5--15 describe the main loop of the algorithm that is executed while the size of the test suite is less than the testing budget. In each iteration of the main loop, the algorithm:

\begin{enumerate}

\item Sets the $target$ variable to the fitness threshold for  an acceptable test candidate. It is initialized to 1, meaning that initially only positive test candidates are considered for inclusion in the test suite. Then, it enters a loop in lines 7--11 that will repeat until a test candidate is found with an outcome equal or greater than the $target$ variable.

\item In the inner loop,  it firsts adjust the $target$ by multiplying by a metaparameter $0<treducer<1$. The goal is to reduce the current target slightly, to make it easier to find a suitable test candidate with every loop iteration.

\item Then, it uses the generator network to create a test candidate that is not already included in the test suite (L9). We use the notation [T]$_1$ to refer to the set of the first elements of the tuples in T. These are the test inputs that have been already executed against the SUT. The set I-[T]$_1$ is therefore the set of test inputs that have not been executed yet.


\item If the predicted fitness of the test candidate is equal or better than the current target (L10) then we exit the inner loop and proceed to execute the test and measure its outcome (L11) and add them to the test suite (L12).

\item Finally, we train the GAN network with the current test suite and iterate the main loop until we have generated all the desired tests.
\end{enumerate}

\subsection{Online GAN Training}

We train the online GAN  following a variation of the traditional GAN training pattern:

First, the discriminator network is trained using the current test suite. The training inputs are all the tests included so far in the test suite and their actual fitness, as established by the SUT during a test execution. In this way, the discriminator is trained to predict the fitness of test candidates. This is a different approach when compared to a traditional GAN, where the discriminator is trained to distinguish between inputs created by the generator and inputs from the training set. It is not possible to apply this approach directly in an online GAN since all the available inputs for the discriminator have been created by the generator. 

Second, once the discriminator is trained, we can proceed to train the whole GAN. For this, we freeze the model parameters for the discriminator and train the generator combined with the already trained discriminator. The objective is to produce from random noise, test candidates with the maximum possible fitness of 1.

Following this pattern, we can train the generator and the discriminator without a prior training set using only the test results obtained online from the SUT, while performing the test generation.



\section{Evaluation}

In this section we present a preliminary evaluation of the online GAN algorithm in an example test system.

\subsection{Task Description}

The goal of the performance exploration for an embedded platform is usually to find in the input space, i.e., from all possible board configurations, the configurations leading to a defined minimum performance level while avoiding, if possible, the ones leading to a high power dissipation level. 

The concrete task is to search for configurations where the dissipated power is greater than 6 W. We set a budget of 200 tests and use three algorithms: the online GAN presented in this article, the Discriminator network DN algorithm presented in~\cite{DBLP:conf/icst/PorresARLT20} and a random search algorithm. The performance of each algorithm is evaluated by its ability to generate tests within the given budget that fulfill the desired performance target (power $\geq$ 6W).

We used as a system under test the workload Core from the EEMBC CoreMark-Pro Suite \footnote{https://github.com/eembc/coremark-pro} running on a ODROID XU3 development board from HARDKERNEL. Its Exynos 5422 processor implements the ARM big.LITTLE architecture with two clusters composed of 4 cores each. The big cluster consists of a high-performance Cortex-A15 quad-core CPUs, and the little cluster is a low power Cortex-A7 quad-core CPUs. 

As input space, we consider the different board configurations in terms of the number of CPUs, type of CPU, core performance level (i.e., Dynamic voltage and frequency scaling, DVFS level) and core utilization level. The core utilization level is set using a scheduler framework, which is based on the concept of setting the CPU bandwidth of a certain process. By setting the runtime of a process inside a period we can control the level of CPU it uses. In this experiment, we considered 10 possible core utilization levels from 10\% to 100\% with 10\% steps.

The system input space has therefore 6 dimensions and is defined for each cluster as the number of CPUs, clock frequency and utilization level. In this experiment, we measure the power dissipation in Watt of the Exynos 5422 processor. There are 479 600 different configurations and executing the benchmark through the entire input space requires around 80 hours. By performing  an exhaustive search we know that 4 736 configurations fulfill the search criteria (1\% of the configurations).


\subsubsection{Feature Modeling and Neural Network architecture}

The architecture of the neural network used in the generator uses three sequential dense layers with 128 neurons each and one output layer. The input layer accepts random noise with 100 dimensions. The output layer has 6 dimensions, corresponding to the test parameters. Each parameter is normalized in the range \([-1,1]\).  The sequential layers use a $tanh$ activation function. The discriminator also uses  three sequential dense layers, with 8 neurons each in this case.  The output layer has only one output, representing the prediction for the performance of the test, in our case the power dissipation, and uses a $Relu$ activation function.

The optimizer used in network training is RMSprop. It is an improvement over Rprop~\cite{DBLP:conf/icnn/RiedmillerB93}, proposed by Geoffrey Hinton and it is expected to perform better when the network is trained in multiple small batches, as it is done by our algorithm.

\subsubsection{Moving Fitness Threshold}

The algorithm uses a moving threshold to decide if it is worth executing a candidate test. This threshold is represented by the variable $target$, and it is initialized to 1, the highest possible value for a fitness function. The threshold is later updated if no test candidates are found by reducing it using a constant $0<treducer<1$.  In the evaluation presented below, we use $treducer=0.95$ with good results.

\subsubsection{ Competing Algorithms}

Finally, we present the two algorithms used as comparison to the new Online GAN algorithm: the Random and the Discriminator network (DN) test generators. 

The Random test generator is probably the simplest test generation algorithm. It  is based on random uniform sampling of the input space $I$. The algorithm is shown in Alg.~\ref{alg3}.
  
\begin{algorithm}
\SetAlgoLined
\textbf{input} set of inputs I, integer budget \\
\textbf{requires} budget $\leq$ |I| \\
T := $\emptyset$\;
 \While{\normalfont{|T|<budget}}{
   t:= uniform(I-[T]$_1$,1); \\
   result := measure(execute(t))\;
   T := T $\cup$ \{(t,result)\}\;
  }
\textbf{result} test suite T
 \caption{Random test generation algorithm}\label{alg3}
\end{algorithm}

The Discriminator network DN algorithm is shown in Alg.~\ref{alg4}. It is based on the algorithm presented in ~\cite{DBLP:conf/icst/PorresARLT20}. The main difference between the GAN and DN algorithms is that the later uses random uniform sampling instead of a GAN to create test candidates, as shown in L9. Since this algorithm relies on sampling of the input space, it generates a number of samples on each loop iteration, as indicated by the $batchsize$ constant, and then it picks the sample with the best predicted fitness. In contrast, the GAN algorithm only generates one test candidate at each iteration.

\begin{algorithm}
\SetAlgoLined
\textbf{input} set of inputs I, integer budget \\
\textbf{requires} budget $\leq$ |I| \\
T := $\emptyset$\;
 \While{\normalfont{|T|<budget}}{
   \Repeat{\normalfont{predict(DN,best) $\geq$ target}}{
   target := target*$treducer$ ; \\
   candidates:= uniform(I-[T]$_1$,$batchsize$); \\
   best := argmax( predict(DN,t), t $\in$ candidates)\;
    }
   result := measure(execute(best))\;
   T := T $\cup$ \{(best,result)\}\;
   train(DN, T)
  }
\textbf{result} test suite T
 \caption{Online DN test generation algorithm}\label{alg4}
\end{algorithm}

\subsection{Experimental Results}

We run each algorithm 10 times, using different random seeds, generating 200 tests in each run. In each run, we first create 50 random tests and then we activate the algorithm under evaluation. We  measure the fitness of each test as the function $\min(1,\frac{power}{6})$. That is, a test yielding a dissipation power greater or equal than 6W has a fitness of 1.

\begin{figure*}
\begin{center}
 \includegraphics[scale=0.4]{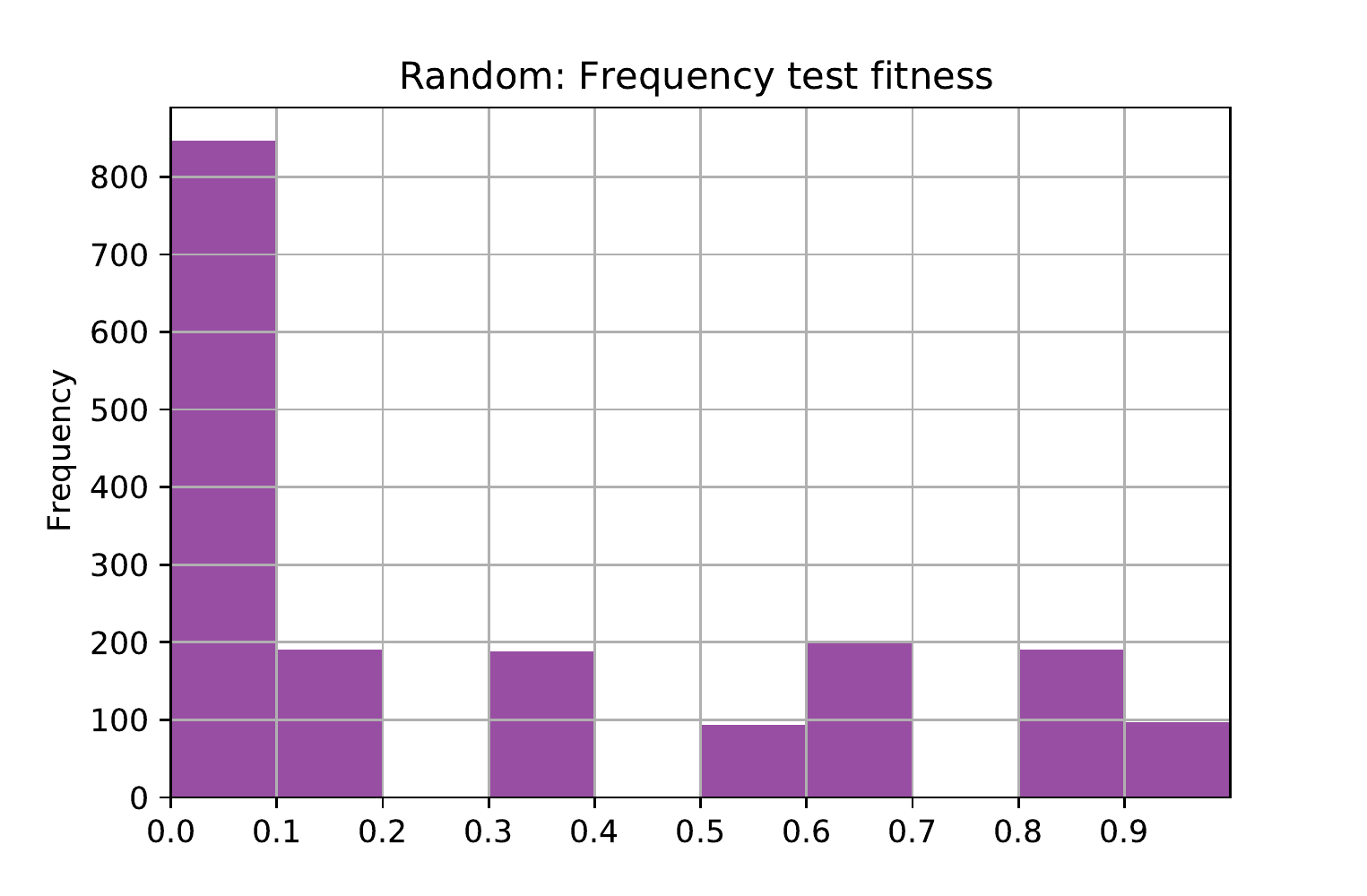} 
  \includegraphics[scale=0.4]{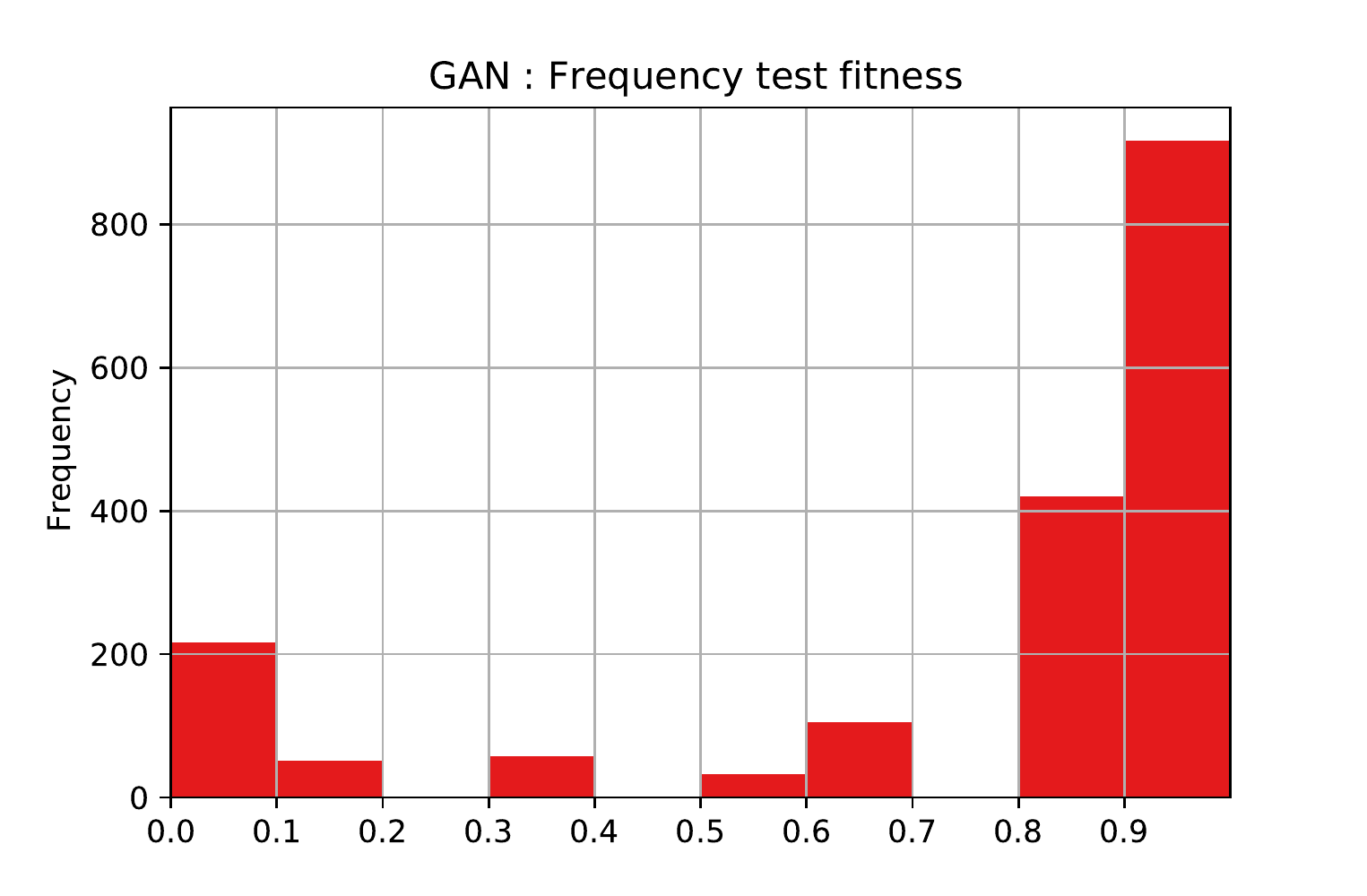} \\
    \includegraphics[scale=0.4]{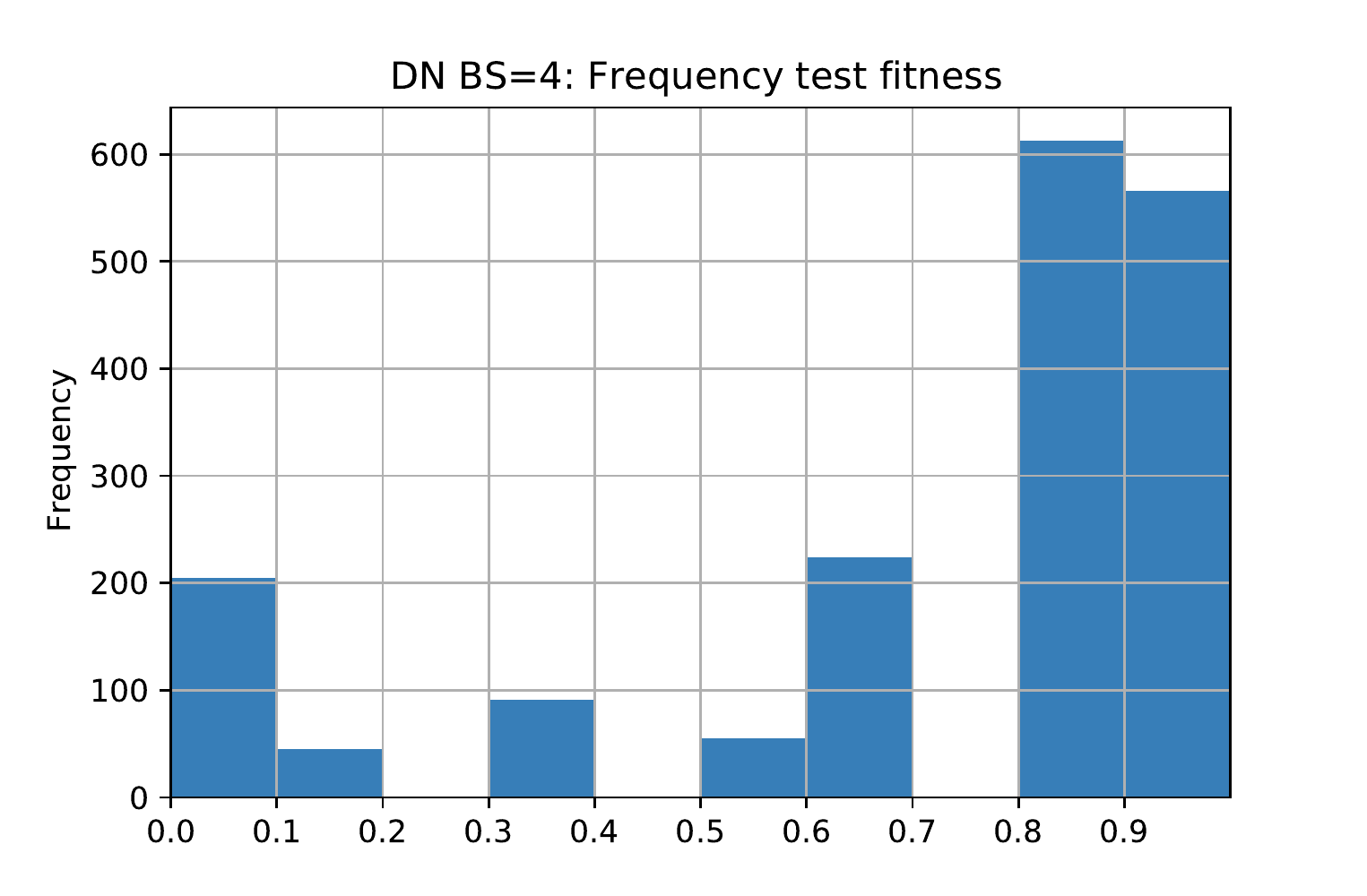} 
      \includegraphics[scale=0.4]{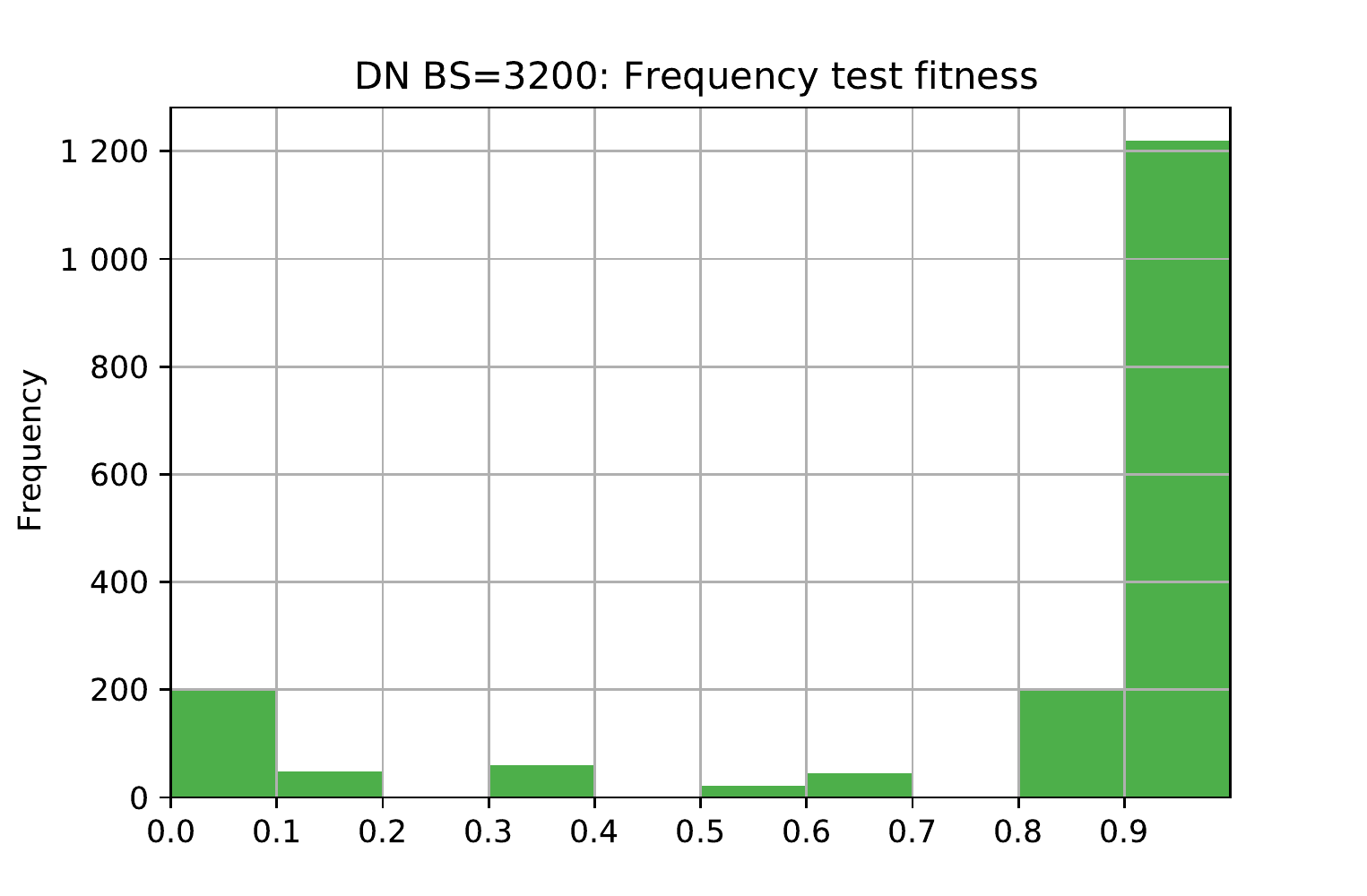}
    \includegraphics[scale=0.5]{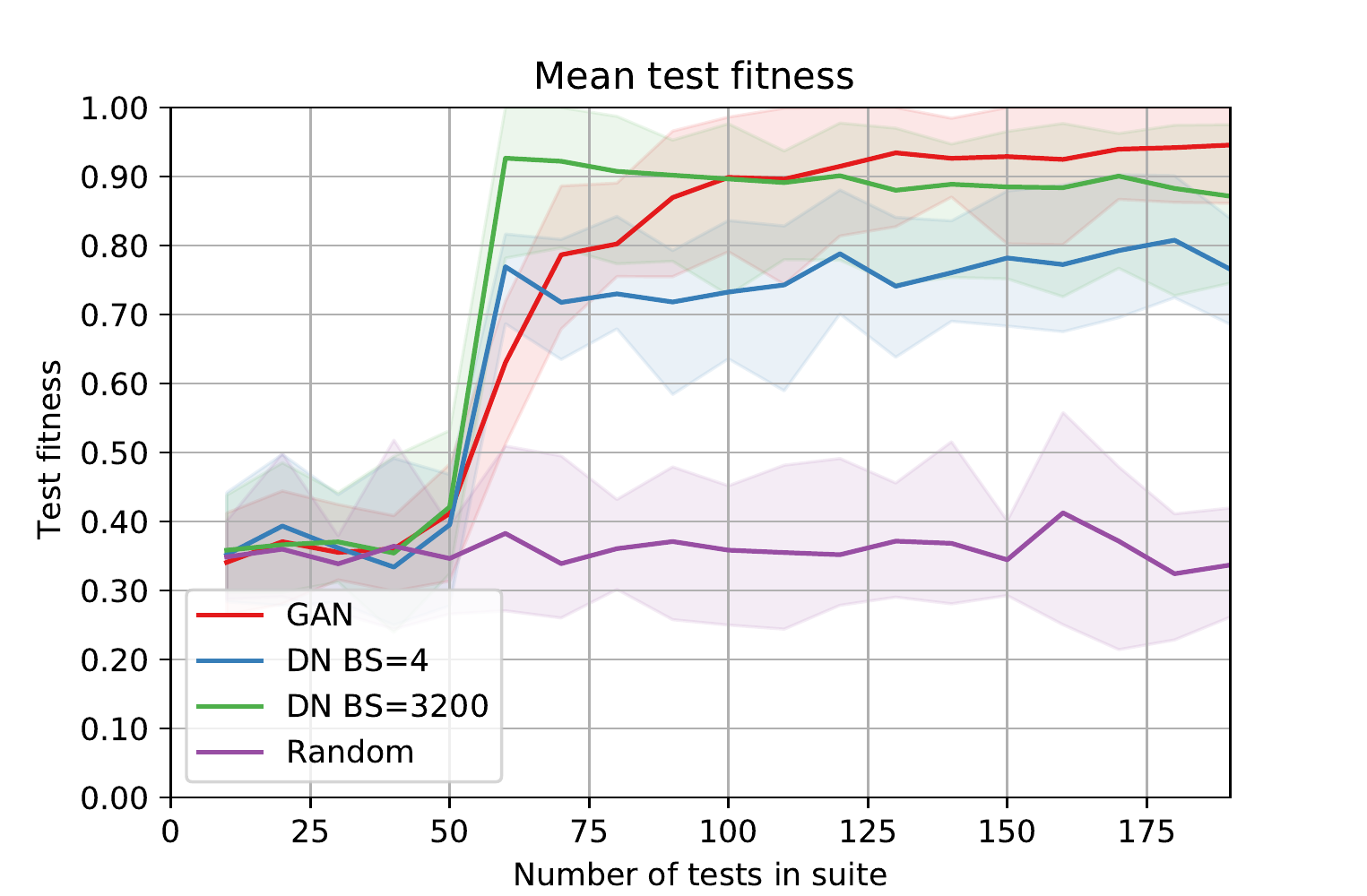} \\
\end{center}
\vspace*{-3mm} 
\caption{Test Fitness for Odroid Power $\geq$ 6W}
\label{fig:tf}
\vspace*{-3mm} 
\end{figure*}

Figure~\ref{fig:tf} shows four histograms for the fitness of the tests generated by each algorithm. The right most column of the histogram represents the number of positive tests that match our search criteria.

We can observe that the random search algorithm generates mostly tests with a  low fitness, as expected. The other algorithms perform better and manage to generate a large number of tests with a high fitness value.

The performance of the DN algorithm, that uses a random generator, depends highly  on the batch size. By changing the batch size the algorithm generates more test candidates until the discriminator selects one for the actual execution, as shown in Table~\ref{t:tries}. The effect of a larger batch size is a better test suite, as we can observe at the two bottom histograms in Figure~\ref{fig:tf}. If we set a small batch size of 4, the algorithm performs on average 6.35 iterations of its inner loop and generates 25.4 candidate tests before one of these tests passes the filter of the discriminator network.  With a batch size of 32 000, the algorithm performs only 4.46 iterations of its inner loop but generates an average of 142 816 test candidates before one of these tests has a high enough fitness.

\begin{table}[]
  \caption{Average Number of Trials per Accepted Test }\label{t:tries}
  \begin{center}
  \begin{tabular}{lll}
  	Algorithm    & Iterations & Trials \\ \hline
  	GAN          & 2.45       & 2.45   \\
  	DN  bs=4     & 6.35       & 25.4   \\
  	DN  bs=32000 & 4.46       & 142816
  \end{tabular}
\end{center}
\end{table}

Compared to the DN algorithm, the new GAN algorithm uses a deep neural network as a generator. This allows it to generate test candidates with high fitness quickly. In our experiment, the algorithm  generates an average of 2.45 test candidates 
per actual test. Still, it produces a test suite populated with a large number of tests with high fitness.

Finally, we can observe in the plot at the bottom of Figure~\ref{fig:tf} the evolution of the simple moving average (SMA) of the mean of the fitness of the test suites generated by each algorithm. The first 50 tests are always generated randomly. After that we can observe that the DN algorithm quickly improves its test suites, converging at a mean that is dependent on its batch size. The online GAN algorithm improves its performance at a slower pace, but after generating 100 additional tests it surpasses the DN algorithm  using a large batch size. We conjecture that is due to the fact that the online GAN algorithm requires more iterations to train its two different networks when compared to the DN algorithm, that uses only one network.

Overall, we consider that the new online GAN algorithm performs competitively when compared to our previous algorithms, creating test suites with a large test fitness without relying on sampling of the solution space.

\section{Conclusions}

\label{sec:concl}

In this paper we present an online GAN algorithm for automatic performance test generation. This algorithm combines training and test generation as well as surrogate model  creation in one single automated step. To our knowledge, this is a novel approach and shows how we can benefit from the abundant research on GAN for the domain of  test generation.

There are many open research problems around this approach.  The current algorithm has several important limitations. It assumes that the SUT is stateless and the outcome of the tests is deterministic. Also, only integer and floating point inputs are supported. We present a preliminary evaluation of the algorithm using only one example system and we acknowledge that a more thorough evaluation is needed.

Still, we consider that the presented algorithm serves as a proof of concept and we hope that it can spark a research discussion on the application of GAN for test generation.

\section*{Acknowledgments}
This research work has received funding from the ECSEL Joint Undertaking (JU) under grant agreement
No 101007350. The JU receives support from the European Union’s Horizon 2020 research and innovation
programme and Sweden, Austria, Czech Republic, Finland, France, Italy, Spain.
 

\bibliography{testing}
\bibliographystyle{plain}
\end{document}